\documentstyle{ichep_pre}
\input epsf
\global\arraycolsep=1.8pt
\def\sss{\scriptscriptstyle}
\def\barp{{\raise.35ex\hbox{${\sss (}$}}---{\raise.35ex\hbox{${\sss )}$}}}
\def\bdbarp{\hbox{$B_d$\kern-1.4em\raise1.4ex\hbox{\barp}}}
\def\bsbarp{\hbox{$B_s$\kern-1.4em\raise1.4ex\hbox{\barp}}}
\newcommand{\xd}{x_d}
\newcommand{\xs}{x_s}

\newcommand{\bs}{B_s^0}
\newcommand{\bsb}{\overline{B_s^0}}

\newcommand{\beq}{\begin{equation}}
\newcommand{\eeq}{\end{equation}}
\newcommand{\absvcb}{\vert V_{cb}\vert}
\newcommand{\absvub}{\vert V_{ub}\vert}

\newcommand{\abseps}{\vert\epsilon\vert}

\newcommand{\fbb}{f^2_{B_d}\hat{B}_{B_d}}
\newcommand{\fbbs}{f^2_{B_s}\hat{B}_{B_s}}
\newcommand{\fbd}{f_{B_d}}

\def\app{p \bar{p}}

\def\mt{m_t}

\newcommand{\delmd}{\Delta M_d}
\newcommand{\delms}{\Delta M_s}

\newcommand{\bdbdbar}{$B_d^0$-${\overline{B_d^0}}$}
\newcommand{\bsbsbar}{$B_s^0$-${\overline{B_s^0}}$}

\begin{document}

\boldmath
\title{An Update of the CKM Matrix$^{+}$}
\unboldmath

\author{Ahmed Ali$^{\dag\ddag\P}$\ and David London$^{\S\|}$}

\affil{\dag\ Theory Division, CERN, CH-1211 Geneva 23, Switzerland\\
\S\ Laboratoire de physique nucl\'eaire, Universit\'e de
Montr\'eal \\
        C.P. 6128, succ. centre-ville, Montr\'eal, QC, Canada}

\abstract{
We update the constraints on the parameters of the quark flavour mixing
matrix $V_{CKM}$ in the standard model using the latest experimental and
theoretical results as input. We present the 95$\%$ C.L. allowed region of
the unitarity triangle and the corresponding ranges for the ratio $\vert
V_{td}/V_{ts} \vert$ and for the quantities $\sin 2\alpha$, $\sin 2\beta$
and $\sin^2\gamma$, which characterize CP-violating rate asymmetries in
$B$-decays. The SM prediction for the $\bs$-$\bsb$ mixing ratio $\xs$ is
also presented.}

\twocolumn[\maketitle]

\fnm{6}{Talk presented by A. Ali at the $27^{th}$ International Conference
on High Energy Physics (ICHEP), Glasgow, Scotland, July 1994}
\fnm{2}{e-mail: alia@cernvm.cern.ch}
\fnm{5}{On leave of absence from DESY, Hamburg, FRG.}
\fnm{4}{e-mail: london@lps.umontreal.ca}

\boldmath
\section{Experimental and Theoretical Input}
\unboldmath

In ref.~\cite{AL94}, we recently updated the profile of the
Cabibbo-Kobayashi-Maskawa (CKM) matrix \cite{CKM}, in particular the CKM
unitarity triangle. In performing this update, we included the improvements
reported in the measurements of the $B$ lifetime, the \bdbdbar\ mass
difference $\delmd$, the CKM matrix elements $\absvcb$ and $\vert
V_{ub}/V_{cb} \vert$ from $B$ decays, measured by the ARGUS, CLEO, CDF and
LEP experiments, and the lower bound on the ratio $\delms/\delmd$ reported
by the ALEPH collaboration \cite{Forty}. The CDF value for the top quark
mass $\mt = 174 \pm 10^{+13}_{-12}$ GeV \cite{CDFmt} was also included. We
refer to ref.~\cite{AL94} for details and references to earlier work and
confine ourselves here to giving the principal results.

In performing this update, we make use of the Wolfenstein parametrization
\cite{Wolfenstein} in which the CKM matrix can be written in terms of four
parameters $\lambda$, $A$, $\rho$ and $\eta$. The matrix element $\vert
V_{us}\vert$ has been extracted from $K\to\pi e\nu$ and hyperon decays to
be $\vert V_{us}\vert=\lambda=0.2205\pm 0.0018$ \cite{PDG}. The parameter
$A$ is related to the CKM matrix element $V_{cb}$, which can be obtained
from semileptonic decays of $B$ mesons. Using methods based on heavy quark
effective theory (HQET), we find
\beq
\absvcb = 0.039 \pm 0.006 ~,
\label{alvcb}
\eeq
which yields $A = 0.80 \pm 0.12$.
We refer to ref.~\cite{AL94} for a full discussion of the experimental and
theoretical inputs leading to the above values.
and to ref.~\cite{HQETICHEP} for the latest developments in HQET.

The parameters $\rho$ and $\eta$ are constrained by the measurements of
$\vert V_{ub}/V_{cb}\vert$, $\abseps$ (the CP-violating parameter in the
kaon system) and $\delmd$ (\bdbdbar\ mixing induced mass difference). The
ratio $\vert V_{ub}/V_{cb}\vert$ can be obtained by looking at the endpoint
of the inclusive lepton spectrum in semileptonic $B$ decays. There still
exists quite a bit of model dependence in the interpretation of data and
the present average is \cite{Casselpc}
\beq
\left\vert \frac{V_{ub}}{V_{cb}} \right\vert = 0.08\pm 0.03~,
\label{vubvcbn}
\eeq
giving $\sqrt{\rho^2 + \eta^2} = 0.36 \pm 0.14$.

Turning to $\abseps$, its experimental value is $\abseps = (2.26\pm
0.02)\times 10^{-3}$ \cite{PDG}. In the SM, $\abseps$ is given by
\begin{eqnarray}
\abseps &=& \frac{G_F^2f_K^2M_KM_W^2}{6\sqrt{2}\pi^2\Delta M_K}
\hat{B}_K A^2\lambda^6\eta
\bigl(y_c\left\{\hat{\eta}_{ct}f_3(y_c,y_t)-\hat{\eta}_{cc}\right\}
 \nonumber \\
&~& ~~~~~~~~~~~~~~+ ~\hat{\eta}_{tt}y_tf_2(y_t)A^2\lambda^4(1-\rho)\bigr).
\label{eps}
\end{eqnarray}
Here, the $\hat{\eta}_i$ are QCD correction factors \cite{QCDcorrs},
$y_i\equiv m_i^2/M_W^2$, and the functions $f_2$ and $f_3$ are given in
ref.~\cite{AL94}.

The final parameter in the expression for $\abseps$ is the renormalization
scale independent parameter $\hat{B}_K$, which represents our ignorance of
the hadronic matrix element $\langle K^0 \vert {({\overline{d}}\gamma^\mu
(1-\gamma_5)s)}^2 \vert {\overline{K^0}}\rangle$. The evaluation of this
matrix element has been the subject of much work, summarized in
ref.~\cite{AL92}. The present lattice-QCD estimates give $\hat{B}_K=0.82 \pm
0.027 \pm 0.023$ \cite{bklattice}. In our fits we take $\hat{B}_K = 0.8 \pm
0.2 $, although we also consider specific values of $\hat{B}_K$ ranging
from 0.4 to 1.

The present world average of the \bdbdbar\ mixing parameter $\xd\equiv
\Delta M_d/\Gamma_d$ is $\xd = 0.76 \pm 0.06$ \cite{Forty}. The precision
on $\Delta M_d$ alone is now quite competitive with the precision on $\xd$.
The LEP-average $\Delta M_d= 0.513 \pm 0.036$ (ps)$^{-1}$ has been combined
with that derived from time-integrated measurements yielding the present
world average \cite{Forty}
\beq
\Delta M_d = 0.500 \pm 0.033 ~\mbox{(ps)}^{-1} ~.
\label{deltamd}
\eeq
In our fits we use this number instead of $\xd$.

\begin{table}
\hfil
\vbox{\offinterlineskip
\halign{&\vrule#&
   \strut\quad#\hfil\quad\cr
\noalign{\hrule}
height2pt&\omit&&\omit&\cr
& Parameter && Value & \cr
height2pt&\omit&&\omit&\cr
\noalign{\hrule}
height2pt&\omit&&\omit&\cr
&  $\lambda$ && $0.2205$ & \cr
&  $\vert V_{cb} \vert $ && $0.039 \pm 0.006$ & \cr
&  $\vert V_{ub} / V_{cb} \vert$  && $0.08 \pm 0.03$ & \cr
&  $\abseps$  && $(2.26 \pm 0.02) \times 10^{-3}$ & \cr
&  $\Delta M_d$ && $(0.50 \pm 0.033)$ (ps)$^{-1}$  & \cr
&  $\overline{\mt}(\mt(pole))$ && $(165 \pm 16)$ GeV & \cr
&  $\hat{\eta}_B$  && $0.55$ & \cr
&  $\hat{\eta}_{cc} $ && $1.10$ & \cr
&  $\hat{\eta}_{ct} $ && $0.36$ & \cr
&  $\hat{\eta}_{tt} $ && $0.57$ & \cr
&  $\hat{B}_K$ && $0.8 \pm 0.2$ & \cr
&  $\hat{B}_B$ && $1.0 \pm 0.2$ & \cr
&  $\fbd$ && $180 \pm 50$ MeV & \cr
height2pt&\omit&&\omit&\cr
\noalign{\hrule}}}
\caption{Parameters used in the CKM fits. Values of the hadronic quantities
$\fbd$, $\hat{B}_{B_d}$ and $\hat{B}_K$ shown are motivated by the lattice
QCD results. In Fit 1, specific values of these hadronic quantities are
chosen, while in Fit 2, they are allowed to vary over the given ranges.}
\label{tabfit}
\end{table}

The mass difference $\Delta M_d$ is calculated from the \bdbdbar\ box
diagram, which is dominated by $t$-quark exchange:
\beq
\label{bdmixing}
\Delta M_d = \frac{G_F^2}{6\pi^2}M_W^2M_B\left(\fbb\right)\hat{\eta}_B y_t
f_2(y_t) \vert V_{td}^*V_{tb}\vert^2~, \label{xd}
\eeq
where $\vert V_{td}^*V_{tb}\vert^2= A^2\lambda^{6}
\left[\left(1-\rho\right)^2+\eta^2\right]$. Here, $\hat{\eta}_B$ is the QCD
correction. In the fits presented here we use the value
$\hat{\eta}_B=0.55$, calculated in the $\overline{MS}$ scheme, following
ref.~\cite{etaB}. Consistency requires that the top quark mass be rescaled
from its pole (mass) value of $\mt =174 \pm 16$ GeV to the value
$\overline{\mt}(\mt(pole))$ in the $\overline{MS}$ scheme, which is
typically about 9 GeV smaller.

For the $B$ system, the hadronic uncertainty is given by $\fbb$, analogous
to $\hat{B}_K$ in the kaon system. In our fits, we take ranges for $\fbd$
and $\hat{B}_{B_d}$ which are compatible with recent lattice QCD results
\cite{Shigemitsu} and QCD sum rule results \cite{Narison}:
\begin{eqnarray}
\fbd &=& 180 \pm 50 ~\mbox{MeV}~, \nonumber \\
\hat{B}_{B_d} &=& 1.0 \pm 0.2 ~.
\label{FBrange}
\end{eqnarray}

Table \ref{tabfit} summarizes all input quantities to our fits.

\section{The Unitarity Triangle}

The allowed region in $\rho$-$\eta$ space can be displayed quite elegantly
using the so-called unitarity triangle. The unitarity of the CKM matrix
leads to the relation $V_{ud} V_{ub}^* + V_{cd} V_{cb}^* + V_{td} V_{tb}^*
= 0$. This can be recast as a triangle relation in the $\rho$-$\eta$ plane,
in which the base of the triangle goes from $(0,0)$ to $(1,0)$, and the
apex is given by the coordinates $(\rho,\eta)$. Thus, allowed values of
$\rho$ and $\eta$ translate into allowed shapes of the unitarity triangle.

In order to find the allowed unitarity triangles, the computer program
MINUIT is used to fit the CKM parameters $A$, $\rho$ and $\eta$ to the
experimental values of $\absvcb$, $\vert V_{ub}/V_{cb}\vert$, $\abseps$ and
$\Delta M_d$. Since $\lambda$ is very well measured, we fix it to its
central value given above. We present here the results from two types of
fits:

Fit 1: Here, only the experimentally measured numbers are used as inputs to
the fit with Gaussian errors; the coupling constants $f_{B_d}
\sqrt{\hat{B}_{B_d}}$ and $\hat{B}_K$ are given fixed values.

Fit 2: Here, both the experimental and theoretical numbers are used as
inputs assuming Gaussian errors for the theoretical quantities. All errors
are combined in quadrature.

We briefly summarize the results of Fit 1. The goal here is to restrict the
allowed range of the parameters ($\rho,\eta)$ for given values of the
coupling constants $f_{B_d} \sqrt{\hat{B}_{B_d}}$ and $\hat{B}_K$. In
ref.~\cite{AL94} we showed that certain values of $\hat{B}_K$ and
$f_{B_d}\sqrt{\hat{B}_{B_d}}$ are disfavoured since they do not provide a
good fit to the data. For example, fixing $\hat{B}_K=1.0$, the fitting
program was used to obtain the minimum $\chi^2$ for various values of
$f_{B_d}\sqrt{\hat{B}_{B_d}}$. The results are given in Table 2, along with
the best fit values of $(\rho,\eta)$. Using $\chi^2_{min}<2.0$ as our
``good fit" criterion (since we have two variables, $\rho$ and $\eta$), we
note that $f_{B_d} \sqrt{\hat{B}_{B_d}} < 120$ MeV and $f_{B_d}
\sqrt{\hat{B}_{B_d}} > 290$ MeV give poor fits to the existing data. We
also note that the $\chi^2$ distribution has two minima, at around $f_{B_d}
\sqrt{\hat{B}_{B_d}} = 160$ and 230 MeV. We do not consider this terribly
significant, since the surrounding values of $f_{B_d} \sqrt{\hat{B}_{B_d}}$
also yield good fits to the data. In addition, we found that, for the lower
value $\hat{B}_K=0.4$, the allowed range of $f_{B_d} \sqrt{\hat{B}_{B_d}}$
is quite restricted, with generally higher values of $\chi^2$ than for the
cases of $\hat{B}_K$ in the range 0.6-1.0. This suggests that the data
disfavour (though do not completely exclude) $\hat{B}_K \leq 0.4$
solutions. Details are given in ref.~\cite{AL94}.

\begin{table}
\hfil
\vbox{\offinterlineskip
\halign{&\vrule#&
   \strut\quad#\hfil\quad\cr
\noalign{\hrule}
height2pt&\omit&&\omit&&\omit&\cr
& $\fbd\sqrt{\hat{B}_{B_d}}$ (MeV)
 && $(\rho,\eta)$ && $\chi^2_{min}$ & \cr
height2pt&\omit&&\omit&&\omit&\cr
\noalign{\hrule}
height2pt&\omit&&\omit&&\omit&\cr
& $110$ && $(-0.48,~0.10)$ && $3.24$ & \cr
& $120$ && $(-0.44,~0.12)$ && $1.77$ & \cr
& $130$ && $(-0.40,~0.15)$ && $0.85$ & \cr
& $140$ && $(-0.36,~0.18)$ && $0.33$ & \cr
& $150$ && $(-0.32,~0.21)$ && $7.6\times 10^{-2}$ & \cr
& $160$ && $(-0.28,~0.24)$ && $1.1\times 10^{-3}$ & \cr
& $170$ && $(-0.23,~0.27)$ && $2.4\times 10^{-2}$ & \cr
& $180$ && $(-0.17,~0.29)$ && $8.0\times 10^{-2}$ & \cr
& $190$ && $(-0.11,~0.32)$ && $0.12$ & \cr
& $200$ && $(-0.04,~0.33)$ && $0.13$ & \cr
& $210$ && $(0.03,~0.33)$ && $8.5\times 10^{-2}$ & \cr
& $220$ && $(0.09,~0.33)$ && $2.8\times 10^{-2}$ & \cr
& $230$ && $(0.15,~0.33)$ && $4.5\times 10^{-5}$ & \cr
& $240$ && $(0.21,~0.33)$ && $4.4\times 10^{-2}$ & \cr
& $250$ && $(0.25,~0.33)$ && $0.18$ & \cr
& $260$ && $(0.29,~0.33)$ && $0.43$ & \cr
& $270$ && $(0.33,~0.33)$ && $0.77$ & \cr
& $280$ && $(0.37,~0.33)$ && $1.21$ & \cr
& $290$ && $(0.40,~0.33)$ && $1.73$ & \cr
& $300$ && $(0.43,~0.32)$ && $2.34$ & \cr
height2pt&\omit&&\omit&&\omit&\cr
\noalign{\hrule}}}
\caption{The ``best values'' of the CKM parameters $(\rho,\eta)$ as a
function of the coupling constant $\fbd\protect\sqrt{\hat{B}_{B_d}}$,
obtained by a minimum $\chi^2$ fit to the experimental data, including the
renormalized value of $m_t=165 \pm 16$ GeV. We fix $\hat{B}_K=1.0$. The
resulting minimum $\chi^2$ values from the MINUIT fits are also given.}
\label{tabbk1}
\end{table}

\begin{figure}
\vskip -0.9truein
\centerline{\epsfxsize 3.0 truein
\epsfbox {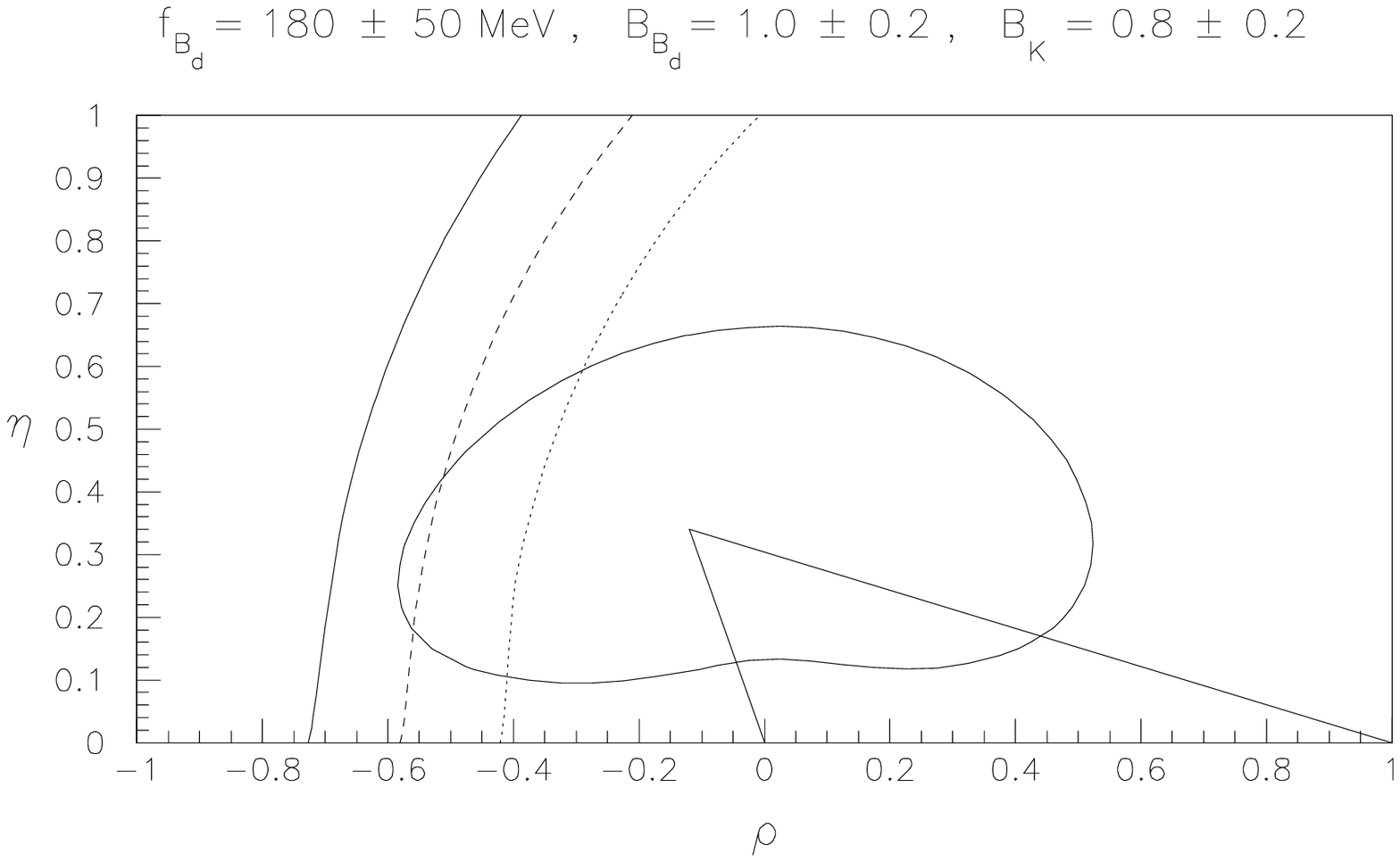}}
\vskip -1.2truein
\caption{Allowed region in $\rho$-$\eta$ space, from a simultaneous fit to
both the experimental and theoretical quantities given in Table
\protect{\ref{tabfit}}. The theoretical errors are treated as Gaussian for
this fit. The solid line represents the 95\% C.L.\ region. The triangle
shows the best fit. The constraints in $\rho$-$\eta$ space from the ALEPH
bound on $\delms$ are presented for 3 choices of the SU(3)-breaking
parameter: $\xi_s^2 = 1.1$ (dotted line), $1.35$ (dashed line) and $1.6$
(solid line). In all cases, the region to the left of the curve is ruled
out.}
\label{xslimit}
\end{figure}

We now discuss Fit 2. Since the coupling constants are not known and the
best we have are estimates given by the ranges in eq.~(\ref{FBrange}), a
reasonable profile of the unitarity triangle at present can be obtained by
letting the coupling constants vary in these ranges. The resulting CKM
triangle region is shown in Fig.~\ref{xslimit}. As is clear from this
figure, the allowed region is enormous! Even so, it is still reduced
compared to the previous such analyses, due to the knowledge of $\mt$.
The preferred values of $\rho$ and $\eta$ obtained from this fit are
\beq
(\rho,\eta) = (-0.12,0.34) ~~~(\mbox{with}~\chi^2 = 1.1\times 10^{-3})~.
\eeq

We have determined bounds on the ratio $\vert V_{td}/V_{ts} \vert$ from our
fits. For $110~\mbox{MeV} \leq f_{B_d} \sqrt{\hat{B}_{B_d}} \leq
290~\mbox{MeV}$, i.e.\ in the entire allowed domain, at 95 \% C.L. we find

\beq
0.11 \leq \left\vert {V_{td} \over V_{ts}} \right\vert \leq 0.36~.
\eeq
The upper bound from our analysis is more restrictive than the current
experimental upper limit following from the CKM-suppressed radiative
penguin decays $BR(B \to \omega + \gamma )$ and $BR(B \to \rho + \gamma )$,
which at present yield $\left\vert {V_{td} /V_{ts}} \right\vert \leq
0.64$--$0.75$ (90\% C.L.) \cite{cleotdul}, depending on the model used for
the SU(3)-breaking in the relevant form factors \cite{SU3ff}. Furthermore,
the upper bound is now as good as that obtained from unitarity, which gives
$0.08 \leq \vert V_{td}/V_{ts} \vert \leq 0.36$, but the lower bound from
our fit is slightly more restrictive.

Note that the matrix element ratio $\vert V_{ub}/V_{cb} \vert$ is very
poorly determined. Our fits give:
\beq
     0.03 \leq  \frac{\absvub}{\absvcb} \leq 0.137 ~.
\label{utsides}
\eeq
It is important to reduce the present errors on the ratios $\vert
V_{td}/V_{ts} \vert$ and $\vert V_{ub}/V_{cb} \vert$ in order to
quantitatively test CKM unitarity.

\section{$\bs$-$\bsb$ Mixing and the Unitarity Triangle}

The \bsbsbar\ box diagram is also dominated by $t$-quark exchange,
and the mass difference between the eigenstates $\delms$ is given by the
analog of eq.~(\ref{xd}),
\beq
\delms = \frac{G_F^2}{6\pi^2}M_W^2M_{B_s}\left(\fbbs\right)
\hat{\eta}_{B_s} y_t f_2(y_t) \vert V_{ts}^*V_{tb}\vert^2~.
\label{xs}
\eeq
A measurement of $\Delta M_s$ can be used to give an additional constraint
on the unitarity triangle. Taking the ratio of $\delmd$ and $\delms$, we
find
\beq
\frac{\delms}{\delmd} =
 \frac{\hat{\eta}_{B_s}M_{B_s}\left(\fbbs\right)}
{\hat{\eta}_{B_d}M_{B_d}\left(\fbb\right)}
\left\vert \frac{V_{ts}}{V_{td}} \right\vert^2.
\label{xratio}
\eeq
All dependence on the $t$-quark mass drops out, leaving the square of the
ratio of CKM matrix elements, multiplied by a factor which reflects
$SU(3)_{\rm flavour}$ breaking effects. Since we expect the QCD correction
factor $\hat{\eta}_{B_s}$ to be equal to its $B_d$ counterpart, the only
real uncertainty in this factor is the ratio of hadronic matrix elements.
In what follows, we take
\beq
\xi_s \equiv {(f_{B_s} \sqrt{\hat{B}_{B_s}}) \over
(f_{B_d} \sqrt{\hat{B}_{B_d}})} = (1.16 \pm 0.1) ~.
\eeq
This is consistent with estimates from lattice QCD \cite{Shigemitsu} and
QCD sum rules \cite{Narison}.

The ALEPH lower bound $\delms /\delmd > 11.3$ at $95\%$ C.L. \cite{Forty}
can thus be turned into a bound on the CKM parameter space $(\rho,\eta)$ by
choosing a value for the SU(3)-breaking parameter $\xi_s^2$. We assume
three representative values: $\xi_s^2 = 1.1$, $1.35$ and $1.6$, and display
the resulting constraints in Fig.~\ref{xslimit}. From this graph we see
that the ALEPH bound marginally  restricts the allowed $\rho$-$\eta$ region
for small values of $\xi_s^2$, but does not provide any useful bounds for
larger values. Of course, an actual measurement of $\delms$ would be very
helpful in further constraining the CKM parameter space.

We now turn to the SM prediction for $\xs\equiv\delms/\Gamma_s$. The main
uncertainty in $\xs$ (or, equivalently, $\delms$) is $\fbbs$. Using the
determination of $A$ given previously, $\tau_{B_s}= 1.54 \pm 0.14$ (ps) and
$\overline{\mt}=165 \pm 16$ GeV, we obtain
\begin{equation}
\xs = \left(19.4 \pm 6.9\right)\frac{\fbbs}{(230~\mbox{MeV})^2}~.
\end{equation}
The choice $f_{B_s}\sqrt{\hat{B}_{B_s}}= 230$ MeV corresponds to the
central value given by the lattice-QCD estimates, and with this our fits
give $\xs \simeq 20$ as the preferred value in the SM.


\section{CP Violation in the $B$ System}

It is expected that the $B$ system will exhibit large CP-violating effects,
characterized by nonzero values of the three angles $\alpha$, $\beta$ and
$\gamma$ in the unitarity triangle. These angles can be measured via
CP-violating asymmetries in hadronic $B$ decays. In the decays $\bdbarp \to
\pi^+ \pi^-$, for example, one measures the quantity $\sin 2\alpha$, and in
$\bdbarp\to J/\psi K_S$, $\sin 2\beta$ is obtained. The CP asymmetry in the
decay $\bsbarp\to D_s^\pm K^\mp$ is slightly different, yielding $\sin^2
\gamma$.

These CP-violating asymmetries can be expressed straightforwardly in terms
of the CKM parameters $\rho$ and $\eta$. The 95\% C.L.\ constraints on
$\rho$ and $\eta$ found previously can be used to predict the ranges of
$\sin 2\alpha$, $\sin 2\beta$ and $\sin^2 \gamma$ allowed in the standard
model. The allowed ranges, obtained from Fit 1, are found in Table
\ref{cpasym1}. In this Table we have assumed that the angle $\beta$ is
measured in $\bdbarp\to J/\Psi K_S$, and have therefore included an extra
minus sign due to the CP of the final state. Since the CP asymmetries all
depend on $\rho$ and $\eta$, the ranges for $\sin 2\alpha$, $\sin 2\beta$
and $\sin^2 \gamma$ shown in Table \ref{cpasym1} are correlated. That is,
not all values in the ranges are allowed simultaneously. This correlation
can be seen in ref.~\cite{AL94}.

\begin{table}
\hfil
\vbox{\offinterlineskip
\halign{&\vrule#&
   \strut\quad#\hfil\quad\cr
\noalign{\hrule}
height2pt&\omit&&\omit&&\omit&&\omit&\cr
& $\fbd\sqrt{\hat{B}_{B_d}}$ && $\sin 2\alpha$ &&
$\sin 2\beta$ && $\sin^2 \gamma$ & \cr
height2pt&\omit&&\omit&&\omit&&\omit&\cr
\noalign{\hrule}
height2pt&\omit&&\omit&&\omit&&\omit&\cr
& $130$ && 0.36 -- 0.96 && 0.17 -- 0.41 && 0.08 -- 0.48 & \cr
& $155$ && 0.15 -- 1.0 && 0.26 -- 0.62 && 0.23 -- 1.0 & \cr
& $180$ && $-$1.0 -- 1.0 && 0.33 -- 0.81 && 0.37 -- 1.0 & \cr
& $205$ && $-$1.0 -- 1.0 && 0.40 -- 0.93 && 0.20 -- 1.0 & \cr
& $230$ && $-$1.0 -- 0.86 && 0.47 -- 0.99 && 0.15 -- 1.0 & \cr
height2pt&\omit&&\omit&&\omit&&\omit&\cr
\noalign{\hrule}}}
\caption{The allowed ranges for the CP asymmetries $\sin 2\alpha$, $\sin
2\beta$ and $\sin^2 \gamma$, corresponding to the constraints on $\rho$ and
$\eta$ obtained in Fit 1. Values of the coupling constant
$\fbd\protect\sqrt{\hat{B}_{B_d}}$ (in MeV) are stated. We fix
$\hat{B}_K=0.8$.}
\label{cpasym1}
\end{table}

Summarizing our results on CP violation, the ranges for the CP-violating
rate asymmetries parametrized by $\sin 2\alpha$, $\sin 2\beta$ and and
$\sin^2 \gamma$ are determined at 95\% C.L. to be
\begin{eqnarray}
&~& -1.0 \leq \sin 2\alpha \le 1.0~, \nonumber \\
&~& 0.17 \leq \sin 2\beta \le 0.99~, \\
&~& 0.08 \leq \sin^2 \gamma \le 1.0~. \nonumber
\end{eqnarray}
(For $\sin 2\alpha < 0.4$, we find $\sin 2\beta \ge 0.3$.)

\end{document}